\documentclass[aip,reprint]{revtex4-1}

\usepackage[utf8]{inputenc} 
\usepackage[T1]{fontenc}    
\usepackage{hyperref}       
\usepackage{url}            
\usepackage{booktabs}       
\usepackage{amsfonts}       
\usepackage{nicefrac}       
\usepackage{microtype}      
\usepackage{graphicx}
\usepackage{lipsum}
\usepackage{comment}
\usepackage[caption=false]{subfig}
\usepackage{textcomp}
\usepackage{soul}           
\usepackage{amsmath}
\usepackage{listings}
\usepackage{chngcntr}
\usepackage{enumitem}
\usepackage[table,xcdraw]{xcolor}
\usepackage{booktabs}
\usepackage{multirow}
\usepackage{hhline}

\definecolor{codegreen}{rgb}{0,0.6,0}
\definecolor{codegray}{rgb}{0.5,0.5,0.5}
\definecolor{codepurple}{rgb}{0.58,0,0.82}
\definecolor{backcolour}{rgb}{0.95,0.95,0.92}

\lstdefinestyle{mystyle}{
    backgroundcolor=\color{backcolour},   
    commentstyle=\color{codegreen},
    keywordstyle=\color{magenta},
    numberstyle=\tiny\color{codegray},
    stringstyle=\color{codepurple},
    basicstyle=\ttfamily\footnotesize,
    breakatwhitespace=false,         
    breaklines=true,                 
    captionpos=b,                    
    keepspaces=true,                 
    numbers=left,                    
    numbersep=5pt,                  
    showspaces=false,                
    showstringspaces=false,
    showtabs=false,                  
    tabsize=2
}
\newcommand{\paragraphtitle}[1]{\textsf{\textbf{\small {#1}}}}

\draft 

\begin{document}

\title[\texttt{mlcolvar}]{A unified framework for machine learning collective variables for enhanced sampling simulations: \texttt{mlcolvar} } 



\author{Luigi Bonati}
\email[]{luigi.bonati@iit.it}
\affiliation{Atomistic Simulations, Italian Institute of Technology, 16156 Genova, Italy}

\author{Enrico Trizio}
\affiliation{Atomistic Simulations, Italian Institute of Technology, 16156 Genova, Italy}
\affiliation{Department of Materials Science, Università di Milano-Bicocca, 20126 Milano Italy}

\author{Andrea Rizzi}
\affiliation{Atomistic Simulations, Italian Institute of Technology, 16156 Genova, Italy}
\affiliation{Computational Biomedicine, Institute of Advanced Simulations IAS-5/Institute for Neuroscience and Medicine
INM-9, Forschungszentrum Jülich GmbH, Jülich 52428, Germany}

\author{Michele Parrinello}
\affiliation{Atomistic Simulations, Italian Institute of Technology, 16156 Genova, Italy}


\date{\today}

\begin{abstract}
    Identifying a reduced set of collective variables is critical for understanding atomistic simulations and accelerating them through enhanced sampling techniques. Recently, several methods have been proposed to learn these variables directly from atomistic data. Depending on the type of data available, the learning process can be framed as dimensionality reduction, classification of metastable states or identification of slow modes. Here we present \verb|mlcolvar|, 
    a Python library that simplifies the construction of  these variables and their use in the context of enhanced sampling through a contributed interface to the PLUMED software.
    The library is organized modularly to facilitate the extension and cross-contamination of these methodologies.
    In this spirit, we developed a general multi-task learning framework in which multiple objective functions and data from different simulations can be combined to improve the collective variables. 
    The library's versatility is demonstrated through simple examples that are prototypical of realistic scenarios. 
\end{abstract}


\maketitle 

Atomistic simulations, and notably molecular dynamics (MD), are powerful tools that act as a computational microscope capable of shedding  light on the mechanism of many physical-chemical processes~\cite{frenkel2001understanding}. 
In recent years machine learning methodologies have had  a significant and transformative role in atomistic simulations. It suffices to mention their use in developing accurate yet computer-efficient interaction potentials~\cite{Behler2007b,Behler2021,Unke2021} or interpreting the simulation results~\cite{Noe2020}.

However, we shall focus here on a different area of impact for machine learning  methods in MD studies, namely that of enhanced sampling~\cite{bonati2021phd,Noe2020,Chen2021,Sidky2020,wang2020machine}. One of MD's well-known limitations is the time scale that standard simulations can cover. Despite much algorithmic and hardware progress, many processes of relevance, like crystallization, chemical reactions,  or protein folding, remain out of the reach of present-day simulation capabilities. This has encouraged the development of methods that can speed up sampling. The vast literature produced in this area bears witness to the relevance of this issue (see, for instance~\cite{valsson2016,henin2022enhanced,yang2019enhanced} and references therein).

Among the many different approaches, we shall focus here on methods that are based on the addition of an external bias potential to the system' Hamiltonian that depends on a small number of functions of the atomic coordinates. These functions are referred to as collective variables (CVs). If the CVs are appropriately chosen,  the bias added will favor transitions between one metastable state and another, eliminating kinetic bottlenecks and speeding up sampling.    
Besides offering a powerful computational tool, the CVs provide  a concise representation that is precious for an understanding of the physical system.

Traditionally, collective variables have been built out of physical-chemical intuition by choosing experimentally measurable quantities (e.g., the distance between the ends of a protein) or directly related to the nature of the process (e.g., distances associated with bonds being formed or broken in the case of a chemical reaction)~\cite{Bussi2015}.
However, by proceeding in this way, it is easy to overlook important slow variables that might  hinder  convergence. Furthermore, the complexity of the problems that can nowadays be simulated  requires a different  and more automatized approach, in which variables are extracted directly from the data of MD simulations. This is the ideal scenario for machine learning (ML) methods, which excel at learning patterns and complex functions directly from data.  
Indeed, data-driven collective variables have been applied to study and accelerate a wide range of physical~\cite{Rogal2019,Karmakar2021,elishav2023collective}, chemical~\cite{Piccini2018,Mendels2018a,Raucci2022,das2023and}, and biological~\cite{bertazzo2021machine,Ansari2022,lamim2020combination,Rizzi2021,badaoui2022combined,sultan2018transferable,ray2023deep} processes, demonstrating the crucial contribution these approaches can bring to atomistic simulations. 
Testing and combining different methods is however difficult due to the lack of a common framework. The available implementations typically support only a single or very few methods in the literature, and the interfaces with the enhanced sampling codes are limited to specific cases~\cite{Chen2021mlcv,Ketkaew2022,trapl2019anncolvar}.

In this manuscript, we present \verb|mlcolvar|, a library written in Python aimed to simplify the construction of data-driven collective variables and their deployment in the context of enhanced sampling through the PLUMED software~\cite{tribello2014plumed}. The goal is to make these methodologies more accessible to the community and to favor new interesting and useful combinations of different approaches. In presenting this work, we also want to offer a unified perspective on recent developments in these methods.

\section{A data-driven approach to collective variables design} 
\label{sec:data-driven-cvs}

In this section, we briefly introduce the problem of data-driven CVs design. We first describe how CVs are represented. Then, we analyze the criteria used for their optimization and connect them to the type of training data required for each of these tasks. Finally, we provide a short perspective on the problem from the point of view of multitask learning. As we will see in sec.~\ref{sec:library}, the relationship between problem and data and the framework provided by multitask learning is strongly reflected in the structure of the library.

\subsection{Collective variables for enhanced sampling}\label{sec:cvs-definition}

Collective variables are formally defined as functions of atomic coordinates: $\mathbf{s} = \mathbf{s}(\mathbf{R})$.
In the context of CV-based enhanced sampling, an external potential $V=V(\mathbf{s})$ which depends on the CVs is added to the system, resulting in an additional force acting on the atoms.
This requires that the CVs are continuous and differentiable functions. Furthermore, CVs are usually designed to be invariant with respect to the symmetries of the system (e.g., translation, rotation, and permutation of identical atoms). 

Learning CVs in a data-driven way implies having a model function parametrized with a set of parameters that need to be optimized on a set of data.
Usually, the training set consists of samples collected from MD simulations, either unbiased or biased through enhanced sampling procedures. However, raw atomic coordinates do not satisfy the symmetries mentioned above. For this reason, they are often pre-processed to obtain a set of input features that provide an invariant representation of the system. In some cases, an alignment~\cite{Hashemian2013} or data-augmentation~\cite{Chen2018b} procedure can be also performed, but this becomes problematic for large systems or in the presence of chemical reactions.
Instead, the choice of input features also provides a way to incorporate physical knowledge into the model. For instance, distances and angles might be used to describe chemical reactions, while for liquid-solid phase transitions, one can employ, for instance, bond order parameters or structure factor peaks~\cite{neha2022collective}. 

Typically, (several) of these input features are combined together to build the CVs via a model function. The functional form of the model typically implies a trade-off between expressiveness and interpretability. Linear models are immediately interpretable but require the identification of a set of essential input features. On the other hand, non-linear models are more expressive, but their interpretation requires additional procedures~\cite{fleetwood2020molecular, jung2023machine, novelli2022characterizing}.
Among them, artificial neural networks (NNs) have become very popular in recent years because they can be used as universal interpolators to represent complex functions of many inputs and many outputs~\cite{bengio2017deep}. They provide a nonlinear transformation of input features through the composition of multiple affine transformations followed by nonlinear activation functions. 
In the context of enhanced sampling, NNs lend themselves well because they provide a continuous and differentiable representation whose derivatives can be computed efficiently by back-propagation, exploiting the automatic differentiation features of ML libraries. Furthermore, they can easily handle a large set of input features. This makes the descriptor choice less critical and at the same time enables scaling to larger systems.


\subsection{Learning approaches}\label{sec:cvs-objectives}

The design of collective variables is often guided by three main objectives:
\begin{enumerate}
    \item Operate a dimensionality reduction of the system to maximize the information enclosed in the latent space (i.e., the space of the CVs)~\cite{ribeiro2018reweighted,Chen2018b,Lemke2019,Hashemian2013}.
    \item Distinguish the different metastable states. This leads us to see the process of CV learning as a supervised learning task, in which the goal is to classify or discriminate states~\cite{Mendels2018,Sultan2018,Bonati2020}.
    \item Reflect the long-term evolution of the system, that is, to describe its slowest modes, which are related to the transitions between long-lived metastable states~\cite{McGibbon2017,Wehmeyer2018b,Schoberl2019a,Wang2019,Mardt2018,Tiwary2016,noe2017collective}. 
\end{enumerate}

In the context of enhanced sampling, the third objective is typically the one we are interested in, but it is not always possible to use it as an operational criterion, e.g., because of the lack of data. In this regard, we often find ourselves in a chicken-and-egg situation~\cite{Chen2018,bonati2021deep}. Extracting efficient CVs requires exploring all the relevant states and transitions between them, but this exploration typically already requires effective CVs. 
Consequently, the first two criteria have been used as surrogate objectives that do not need dynamical data. Moreover, iterative approaches can be applied to refine the CVs when new data become available, either by performing multiple iterations of the same method or by enforcing more CVs objectives~\cite{Chen2018,bonati2021deep,Belkacemi2022,Chen2023}.

\begin{figure*}
    \centering
    \includegraphics[width=0.7\linewidth]{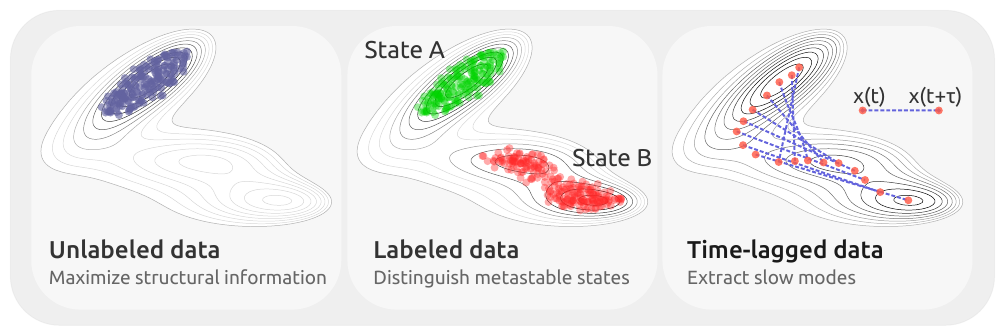}
    \caption{Type of data and the respective optimization criteria which can be used to design data-driven collective variables. 
    }
    \label{fig:data}
\end{figure*}

Each of the objectives usually requires different types of data, as summarized in fig.~\ref{fig:data}. Below, we discuss three typical learning scenarios from this perspective. 

\subsubsection{Working with unlabeled data: unsupervised learning}

The first scenario is the one in which we only have a collection of samples from MD configurations. In this case, we can use unsupervised machine learning techniques, whose goal is to automatically find structural patterns within the data~\cite{Tribello2019,Ceriotti2019}. These methods aim to satisfy the first objective of the CVs in sec.~\ref{sec:cvs-objectives}. Their advantage is that they are applicable to any kind of MD simulations, including those out-of-equilibrium or in which it is not feasible to recover the unbiased dynamics. 
In the context of atomistic simulations, unsupervised learning has been used to identify CVs based solely on a set of configurations from MD simulations. Two notable methods in this category are principal component analysis (PCA), which finds the directions of maximum variance, and autoencoding (AE) NNs~\cite{Chen2018,ribeiro2018reweighted}, which learn a compressed representation with the constraint of being still able to reconstruct the original data. 

These methods are often employed iteratively, alternating cycles of enhanced sampling (biasing along the CV to obtain new data) and CV discovery (optimizing the model on new data, possibly after reweighting)~\cite{Chen2018,ribeiro2018reweighted,Belkacemi2022}.
A typical application of this workflow is the discovery of metastable states. For example, one might start from some reactant molecules and iteratively search for possible products. 


\subsubsection{Classifying metastable states: supervised learning}

A second scenario is the one in which we are aware of the metastable states of interest. These could be, for instance, the reactants and products of a chemical reaction, the folded/unfolded state of a protein, a ligand inside/outside the binding site, or different phases in a material. 
In this setting, we can collect labeled data by performing a short unbiased MD simulation in each state. In a rare event setting, the system will remain trapped in the metastable state thus, the configurations can be easily labeled according to the corresponding state.

This data will allow us not only to perform a dimensionality reduction but also to optimize the CVs to separate the metastable states. This corresponds to the second objective in sec.~\ref{sec:cvs-objectives}.
Among the linear methods, support vector machines~\cite{Sultan2018} as well as Fisher's linear discriminant analysis (LDA)~\cite{Mendels2018} have been employed. Non-linear generalizations based on NNs (Deep-LDA~\cite{Bonati2020}, Deep-TDA~\cite{trizio2021enhanced}) have also been applied to a wide range of physical systems~\cite{Rizzi2021,Karmakar2021,Ansari2022}. This concept has also been applied to build CVs able to drive phase transitions starting from NNs optimized to classify local environments~\cite{Rogal2019}.

Training CVs in a supervised manner provides an easy way of inserting previous knowledge into the CV, either in the form of state classification or also regression of physical observables.
In addition, the resulting CVs can be seen as an initial hypothesis for sampling reactive trajectories between the known states, which can later be used to further refine the CVs. 

\subsubsection{Extracting the slow modes: time-informed learning}

The third setting is the one in which we have reactive simulations that make transitions between different metastable states. In biophysics, these simulations might come from long unbiased simulations performed with purpose-built supercomputers~\cite{lindorff2011fast}, or by changing thermodynamic parameters (e.g. temperature). However, the free energy barriers are often so high that these processes cannot be observed directly. This is particularly the case for chemical reactions and phase transitions. So the most common source of these reactive trajectories is enhanced sampling simulations. 
They can be performed using CVs derived from physical intuition as well as data-driven CVs optimized as discussed above or also via CV-independent methods~\cite{bonati2021deep}.
However, it should be noted that it is not straightforward to recover the unbiased dynamics from biased simulations, although several approximations have been proposed (see, e.g., ~\cite{Chen2023} for a discussion).

Time-informed learning approaches attempt to directly estimate the  modes that govern the long-time evolution of the system (objective 3 in sec.~\ref{sec:cvs-objectives}). In a rare event scenario, these slow modes are indeed related to the transitions between long-lived metastable states. An example of this family is time-lagged independent component analysis (TICA)~\cite{molgedey1994separation,Naritomi2011,Perez-Hernandez2013}, which seeks for the linear combination of the input data that is maximally autocorrelated. Several non-linear generalizations have been proposed to represent better the slow modes~\cite{Mardt2018,Chen2019,bonati2021deep}. Another related class of methods is based on time-lagged autoencoders~\cite{Wehmeyer2018b,Hernandez2018}, which seek to learn a compressed representation from which future configurations can be predicted.


\subsection{Multi-task learning for CV design}

In the previous discussion, constructing a CV was related to the optimization of a specific loss function. However, multiple objectives can be, in principle, combined. In the ML field, this practice is referred to as multi-task learning~\cite{ruder2017overview} and is often adopted to improve the models' generalization capability. Here we discuss how this path can be followed for the CV design and how it motivates the need for a unified framework.


\subsubsection{A multi-task learning perspective on data-driven CVs}

By multi-task learning, we refer to a broad set of algorithms that optimize a ML model based on multiple related objectives.
This concept has been explicitly applied to CVs design in the context of supervised learning~\cite{sun2022multitask}. Moreover, some of the data-driven CVs in the literature can be seen as optimizing multiple objectives. Indeed, NNs that are trained on linear combinations of loss functions also belong to this family. To illustrate this, we refer to two examples. In the case of the EncoderMap method~\cite{Lemke2019}, the reconstruction and the Sketch-map losses are simultaneously optimized, i.e. $\mathcal{L}=\ell_{reconstruction}+\ell_{Sketch-map}
$.
The aim of the Sketch-map~\cite{Ceriotti2011} loss is to encode more structure into the learned CVs by enforcing the distances between points in the latent space to be similar to the corresponding distances in the high-dimensional input space.
Similarly, in the Variational Dynamics Encoder method~\cite{Hernandez2018}, a time-lagged variational autoencoder is optimized together with a cost function that maximizes the autocorrelation of the CV, i.e. $\mathcal{L}=\ell_{reconstruction}+\ell_{autocorrelation}$.


\subsubsection{Learning multiple tasks on different datasets}

In the context of CVs optimization, multi-task learning is usually applied to a dataset with data of the same type. However, we often have different types of data, e.g. labeled and unlabeled, which carry different information that we would like to encode in a single CV. 
To this end, we can resort to a multi-task learning procedure in which a single CV model is optimized using a linear combination of multiple loss functions evaluated on the different datasets. 


This framework provides a straightforward way to combine the different CV objectives discussed in sec.~\ref{sec:data-driven-cvs} to develop new solutions. As an example, in sec.~\ref{sec:examples}, we present a semi-supervised approach obtained by combining different datasets.


\vspace{1em}
To summarize, in this first section, we described the problem of CV construction with a perspective focused on two aspects: the different tasks depending on the type of data available and the combination of models and objective functions for improving CVs.
These considerations have guided the development of the library, which we now illustrate in the next section.

\section{The \texttt{mlcolvar} library}\label{sec:library}

\verb|mlcolvar|, short for Machine Learning COLlective VARiables, is a Python library aimed to help design data-driven CVs for atomistic simulations. The guiding principles of \verb|mlcolvar| are twofold. On the one hand, to have a unified framework to help test and utilize (some of) the CVs proposed in the literature. On the other, to have a modular interface that simplifies the development of new approaches and the contamination between them.

The library is based on the PyTorch machine learning library~\cite{paszke2019pytorch}, and the high-level Lightning package~\cite{falcon2023lightning}. The latter simplifies the overall model training workflow and allows focusing only on the CV design and optimization. 
Although the library can be used as a stand-alone tool (e.g., for analysis of MD simulations), the main purpose is to create variables that can be used in combination with enhanced sampling methods through PLUMED C++ software~\cite{tribello2014plumed} (see sec.~\ref{subsec:deploying}). Hence we will need to deploy the optimized model in a Python-independent and transferable format in order to use it during the MD simulations. 

\begin{figure*}[t!]
    \includegraphics[width=0.85\textwidth]{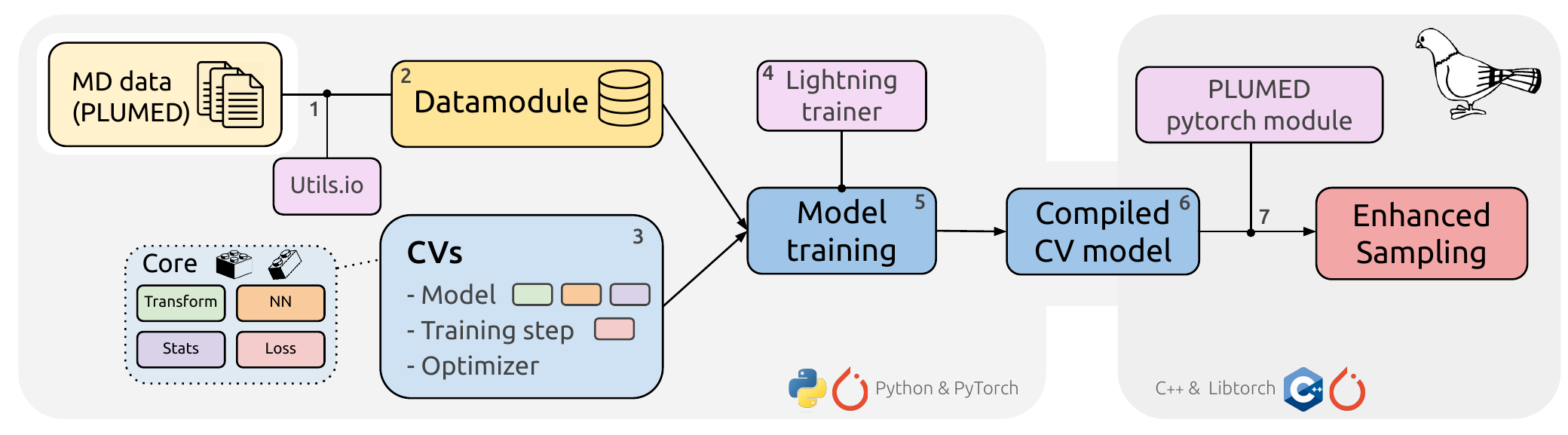}
    \caption{Schematic summary of the workflow for the construction of data-driven CVs in \texttt{mlcolvar}. Data from MD simulations are loaded and framed in a datamodule. A CV is selected from ready-to-use ones (\texttt{mlcolvar.cvs}) or built from the implemented building blocks (\texttt{mlcolvar.core}). After training, the model is compiled with the TorchScript language to be deployed to PLUMED for using it as CV to enhance sampling. A corresponding example of code is given in list.~\ref{lst:workflow}. }
    \label{fig:workflow}
\end{figure*}

\begin{figure*}[t!]
\begin{lstlisting}[language=Python, caption={Example of the typical workflow of CVs optimization with \texttt{mlcolvar}, as schematically depicted in Fig.~\ref{fig:workflow}. The input features are calculated by PLUMED and the end result is a serialized model that can be deployed in PLUMED via the LibTorch C++ interface.}, label=lst:workflow]
# Setup
import torch,lightning
from mlcolvar.data import DictModule
from mlcolvar.cvs import AutoEncoderCV
from mlcolvar.utils.io import create_dataset_from_files

# 1. Import training data (e.g. PLUMED COLVAR files)
dataset = create_dataset_from_files('./COLVAR')
# 2. Create a Lightning datamodule which splits dataset in train/valid
datamodule = DictModule(dataset, lenghts=[0.8,0.2])
# 3. Choose a model and hyper-parameters
cv_model = AutoEncoderCV(encoder_layers=[45,30,15,2])
# 4. Define a trainer object
trainer = lightning.Trainer(max_epochs=1000)
# 5. Optimize parameters
trainer.fit(cv_model, datamodule)
# 6. Compile the model with TorchScript
cv_model.to_torchscript('model.ptc')
# 7. Use it in PLUMED via the pytorch module
\end{lstlisting}
\end{figure*}

\subsection{The CVs optimization workflow}

Here we present the typical workflow used for CVs optimization with \verb|mlcolvar|, from the raw data to an optimized model ready to be used in MD simulations. It is composed of only a few steps which are schematically depicted in Fig.~\ref{fig:workflow}, and a working example of the corresponding few lines of code is reported in Listing~\ref{lst:workflow}. To set the context, we will first provide a practical overview of this general workflow, before delving into the technical details of the individual components in the next sections.

The starting point of our workflow is the data collected during MD simulations with the help of PLUMED. Such data are imported (step 1) in Python and framed in a \verb|DictModule| object (step 2), which optionally divides the data into training and validation datasets (e.g., for early stopping or hyperparameter searching~\cite{bengio2017deep}) and sets up the mini-batches for the training (step 3, see also sec.~\ref{subsec:high-level}). 
The second major ingredient is the \verb|cv_model| to be optimized. In most cases, this is initialized as one of the ready-to-use CV classes already implemented in the library (see sec.~\ref{sec:CVs}). Alternatively, the model class can also be implemented by the user starting from the \verb|core| building blocks provided in the library (see sec.~\ref{subsec:high-level}).
The \verb|cv_model| encapsulates the trainable parameters of the model, the loss function, and the optimizer used for the training.
The optimization is conveniently performed by means of a Lightning \verb|Trainer| object (step 4). In addition to the optimization task, Lightning takes care of many ancillary features, such as applying early stopping, storing metrics or generating log files and checkpoints, and automatically moving data and models to devices such as GPUs. 
After having optimized the \verb|cv_model| on the data, we can deploy it via TorchScript language to make it accessible for production (step 6). At this point, the optimized model can be imported in PLUMED using the \verb|pytorch| model interface (see sec.~\ref{subsec:deploying}) and used as CV for enhanced sampling simulations (step 7).

This minimal workflow can be implemented with \verb|mlcolvar| in 6 lines of code.
The defaults in the library are chosen to offer reasonable starting points.
Nevertheless, while simplifying considerably the training of CVs, the classes are easily customizable and extensible to control aspects related to, for instance, input standardization, network architecture, loss functions, and training hyperparameters.
The technical details of these customizations are discussed at length in the documentation and the tutorials available online (see Code and Data Availability section).

\subsection{High-level overview of the code }\label{subsec:high-level}

In the following, we briefly describe the structure of the \verb|mlcolvar| library and its main modules, which are \verb|data|, \verb|core|, \verb|cvs|, and \verb|utils|.

In \verb|mlcolvar.data|, we provide PyTorch- and Lightning-compatible classes that simplify and improve the efficiency of data access. The key elements are:
\begin{itemize}
    \setlength\itemsep{0em}
    \item \verb|DictDataset|: A dictionary-like PyTorch \verb|Dataset| that maps keys (e.g., data, labels, targets, weights) to tensors. 
    \item \verb|DictLoader|: A PyTorch DataLoader that wraps a \verb|DictDataset|. This class is optimized to significantly reduce the data access time during training and to combine multiple datasets for multi-task training.
    \item  \verb|DictModule|: A Lightning \verb|DataModule| that takes care of automatically splitting a \verb|DictDataset| into training and validation (and optionally test) sets and returning the corresponding dataloaders.
\end{itemize}

In \verb|mlcolvar.core| we implemented the building blocks that are used for the construction of the CV classes. We organized them into the following submodules:

\begin{itemize}
    \setlength\itemsep{0em}
    \item \verb|nn|: learnable modules (e.g., neural networks).
    \item \verb|loss|: loss functions for the CVs optimization.
    \item \verb|stats|: statistical analysis methods (e.g., PCA, LDA, TICA).
    \item \verb|transform|: non-learnable transformations of data (e.g., normalization).
\end{itemize}
All of them are implemented as Python classes that inherit from \verb|torch.nn.Module|. In particular, the \verb|mlcolvar.nn| module contains a class that constructs a general feed-forward neural network which can be customized in several aspects, such as activation functions, dropout, and batch-normalization.

The \verb|mlcolvar.cvs| module includes ready-to-use CV classes, grouped by the type of data used for their optimization in the following sub-packages: 
\begin{itemize}
    \setlength\itemsep{0em}
    \item \verb|unsupervised|: methods that require input data characterizing single MD snapshots.
    \item \verb|supervised|: require labels of the data (e.g., the metastable states they belong to) or a target to be matched in a regression task.
    \item \verb|timelagged|: require pairs of time-lagged configurations, typically from reactive trajectories, to extract the slow modes.
\end{itemize}
Since they are the key element of this library, the structure of CVs is described in more detail in the next subsection. 

Finally, in \verb|mlcolvar.utils| one can find a set of miscellaneous tools for a smoother workflow.
For example, we implemented here helper functions to create datasets from text files, as well as to compute free energy profiles along the CVs. 

\subsection{The structure of CV models}
These CVs are defined as classes that inherit from a \verb|BaseCV| class and \verb|LightningModule|. The former defines a template for all the CVs along with common helper functions, including the handling of data pre- and post-processing. The latter is a Lightning class which adds several functionalities to simplify training and exporting the CV. In particular, \verb|LightningModule| not only encapsulates the model with its architecture and parameters but also implements the training step (and hence defines the loss function) as well as the optimization method. 

The structure of the CVs in \verb|mlcolvar| is designed to be modular. The core of each model is defined as a series of building blocks (typically implemented in \verb|mlcolvar.core|) that are by default executed sequentially, although this can easily be changed by overloading the forward functions in \verb|BaseCV|. An example where this is necessary are AutoEncoders-based CVs. In this case, the building blocks are normalization, encoder and decoder, but the CV is the output of the encoder, not the final output of all the blocks.
New CVs require implementing a \verb|training_step| method that contains the steps that are executed at each iteration of the optimization. Moreover, the loss function and the optimizer settings are saved as class members to allow for easy customization.

In addition, it is possible to add preprocessing and postprocessing layers. 
This allows to speed up the training by applying the transformations only once to the dataset and later including them in the final model for production.
In addition, it allows post-processing to be performed on the model after the training stage (e.g., standardizing the outputs).

Finally, multi-task learning is supported through the \verb|MultiTaskCV| class that takes as input a given model CV as the main task together with a list of (auxiliary) loss functions which will be evaluated on a list of datasets. 
The samples from different datasets go through the same network but enter only one of the loss functions (see panel d of Table~\ref{tab:methods}). The loss function used in this case is a linear combination of each specific loss. Each loss function can optionally be preceded by task-specific layers that are also optimized during the training but are not evaluated to compute the CV. Examples of task-specific models are the decoder used for the reconstruction task in autoencoders (see fig.~\ref{fig:multitask} for an example) or a classifier/regressor used for supervised tasks~\cite{sun2022multitask}. 


\subsection{Deploying the CVs in PLUMED}\label{subsec:deploying}

Once optimized, the CVs are exported using just-in-time compilation to the TorchScript language, returning a Python-independent and transferable model. We wrote an interface within the PLUMED software that allows these exported models to be loaded through the LibTorch library (PyTorch C++ APIs). This is implemented in the \verb|pytorch| module of PLUMED as a function that takes as input a set of descriptors and returns the CVs alongside their derivatives with respect to the input. An example of a minimal input file is shown in Listing~\ref{lst:plumed}. 

This means that the CVs can be immediately used in combination with all enhanced sampling methods implemented in PLUMED (among which we find, for example, Umbrella Sampling~\cite{torrie1977nonphysical}, Metadynamics~\cite{laio2002escaping} and its many variants~\cite{bussi2020using}, Variationally Enhanced Sampling~\cite{valsson2014variational,bonati2019neural}, On-the-fly Probability Enhanced Sampling~\cite{invernizzi2020rethinking}, just to name a few) and within the supported molecular dynamics codes (including but not limited to LAMMPS, GROMACS, AMBER, CP2k, QUANTUM ESPRESSO, and ASE), which allows simulating a wide range of complex processes in (bio)physics, chemistry, materials science, and more.

Note that the interface is very general and thus can be used not only to compute collective variables but also to test CVs defined through complex functions by taking advantage of PyTorch's automatic differentiation capabilities (in the spirit of the PYCV PLUMED module~\cite{Giorgino2019} based on Jax). For example, it has been used to construct CVs from the eigenvalues of the adjacency matrix~\cite{Raucci2022}.

\begin{figure*}[t]
\begin{lstlisting}[language=Python, caption={Example of PLUMED input file in which the calculation of input features is requested, the CV model exported from \texttt{mlcolvar} is loaded, and an enhanced sampling calculation is performed on it.},label=lst:plumed]
# 1. Compute input features (e.g. pairwise distances)
d1: DISTANCE ATOMS=1,2
...
dN: DISTANCE ATOMS=17,19
# 2. Load model	exported by mlcolvar
cv: PYTORCH_MODEL FILE=model.ptc ARG=d1,...,dN
# 3. Apply bias potential (e.g. with OPES)
opes: OPES_METAD ARG=cv.node-0 PACE=500 BARRIER=40
\end{lstlisting}
\end{figure*}

\subsection{Code dependencies}

We kept the number of library dependencies as small as possible. In particular, PyTorch~\cite{paszke2019pytorch}, Lightning~\cite{falcon2023lightning}, and NumPy~\cite{harris2020array} are required, while Pandas~\cite{reback2020pandas} is recommended to efficiently load data from files. 

In order to use the CVs in PLUMED, we need to configure it with the LibTorch C++ library and enable the PLUMED \verb|pytorch| module.

\section{Methods for CVs optimization} \label{sec:CVs}
In this section, we briefly present the methods implemented in the \verb|mlcolvar| library for the identification of collective variables. Furthermore, we will mention some related methods and show how they can be constructed based on the building blocks of the library. In Table ~\ref{tab:methods} we summarized them together with an overview of the common architectures. Note that this does not want to be an extensive review or comparison of the different methods, but rather a concise reference describing the different approaches that can be employed in a given scenario. 
In each section, we start discussing a linear statistical method and then move to the neural-network CVs which are implemented in \verb|mlcolvar|.

\begin{table*}
\centering
\begin{tabular}{lllll} 
\hline\hline
\textbf{Data}                          & \textbf{Objective}                                                                                                         & \textbf{Method}                    & \textbf{Architecture} & \textbf{Notes}                                                                     \\ 
\hline\hline
\multirow{4}{*}{\textbf{Unlabeled }}   & \multirow{4}{*}{\begin{tabular}[c]{@{}l@{}}\textit{Maximize }\\\textit{structural }\\\textit{information}\end{tabular}}    & PCA                                & linear                                                                           &                                                                                   \\
                                       &                                                                                                                            & AutoEncoder (AE)                   & \textbf{A}~                                                                               &                                                                                   \\
                                       &                                                                                                                            & Variational AE (VAE)               & \textbf{A}~                                                                               &                                                                                   \\
                                       &                                                                                                                            & \textit{EncoderMap}                         & \textbf{A}~                                                                               & AE+Sketch-Map loss                                                                \\ 
\hline
\multirow{3}{*}{\textbf{Labeled }}     & \multirow{3}{*}{\begin{tabular}[c]{@{}l@{}}\textit{Distinguish }\\\textit{metastable }\\\textit{states}\end{tabular}}      & LDA                                & linear                                                                           &                                                                                   \\
                                       &                                                                                                                            & Deep-LDA                           & \textbf{B}                                                                                &                                                                                   \\
                                       &                                                                                                                            & Deep-TDA                           & \textbf{C}                                                                                &                                                                                   \\ 
\hline
\multirow{4}{*}{\textbf{Time-lagged }} & \multirow{2}{*}{\textit{Slow modes}}                                                                                        & TICA                               & linear                                                                           &                                                                                   \\
                                       &                                                                                                                            & Deep-TICA/SRV                      & \textbf{B}                                                                                &                                                                                   \\ 
\cline{2-5}
                                       & \multirow{2}{*}{\begin{tabular}[c]{@{}l@{}}\textit{Slow modes}\\\textit{+ structural }\\\textit{information}\end{tabular}} & \textit{Time-lagged AE (TAE)}               & \textbf{A}                                                                                & AE with time-lagged dataset                                                       \\
                                       &                                                                                                                            & \textit{Variational Dynamics
                                       Encoder (VDE)} & \textbf{A}                                                                                & \begin{tabular}[c]{@{}l@{}}Time-lagged VAE \\+ autocorrelation loss\end{tabular}  \\
\hline\hline
\end{tabular}
\vspace{0.5em}

\includegraphics[width= \linewidth]{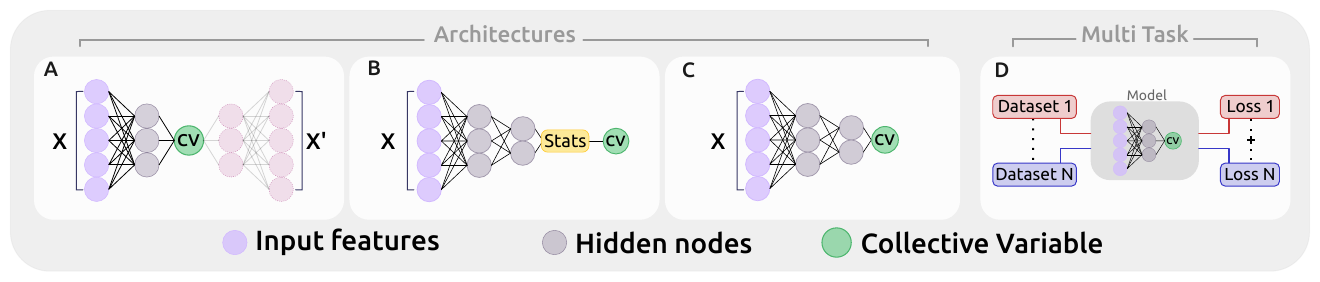}
\caption{(top) List of the methods implemented in \texttt{mlcolvar} for building CVs grouped accordingly to the data used for their optimization. In italics the method which can be easily implemented by following the related notes. (bottom) Sketch of the different neural network CV architectures. a) AutoEncoder, composed by an encoder that maps the input data to a latent space (the CVs) and a decoder that reconstructs it. b) Feed-forward NN used to transform the inputs before applying a statistical method (e.g., LDA, TICA). The CVs are obtained as linear combinations of the NN outputs (e.g. DeepLDA, SRV/DeepTICA).
    c) Feed-forward NN whose outputs are used directly as collective variables (e.g. DeepTDA).}
    \label{tab:methods}
\end{table*}

\subsection{Unsupervised methods}

\vspace{0.5em}
\paragraphtitle{Principal Component Analysis (PCA)}~\cite{jolliffe2002principal} is a linear dimensionality reduction technique that projects the data on the principal components, i.e., the directions of maximum variance. These directions correspond to the eigenvectors of the data covariance matrix, while its eigenvalues measure the amount of explained variance. PCA is typically used to process inputs and provide whitened descriptors for other models or even directly as CVs, in which case only the very first few components are used.

\vspace{0.5em}
\paragraphtitle{AutoEncoders (AEs)} are a class of NNs consisting of two main components: an encoder and a decoder (see panel a in Table ~\ref{tab:methods}). The encoder maps the input descriptors into a low-dimensional latent space, while the decoder performs the inverse task, i.e., reconstructing the original input from its low-dimensional representation.
AEs are trained by minimizing the reconstruction loss, which is usually measured as the mean square error (MSE) between the input and its reconstructed output. The latent space thus learns a minimal set of features that maximally preserves the information on the input structures, and, in this sense, AEs can be viewed as a non-linear generalization of PCA. During the simulation, the output of the encoder is used as CV, while the decoder is used only during training~\cite{Chen2018,Belkacemi2022}. 

\vspace{0.5em}
\paragraphtitle{Variational AutoEncoders (VAEs)}~\cite{kingma2013auto} are a probabilistic variant of AEs, which mainly differ from standard autoencoders in that the data in the latent space is pushed to follow a predefined prior distribution, which is normally a Gaussian distribution with zero mean and unit variance, $\mathcal{N}(0,1)$.
This acts as a regularizer and encourages the network to learn a continuous and smoother latent space representation.
This is accomplished by modifying both the network architecture and the loss function. First, the encoder learns to output the mean and variance of a Gaussian distribution, and the sample that goes through the decoder is drawn from this Gaussian.
Second, the encoder/decoder parameters are optimized to minimize a linear combination of the reconstruction loss and the Kullback-Leibler (KL) divergence between the Gaussian learned by the encoder and the prior distribution $\mathcal{N}(0,1)$.
As CV, the implementation in the \verb|mlcolvar| library then uses only the output of the encoder corresponding to the mean (i.e., ignoring the variance output). 

\vspace{0.5em}
\paragraphtitle{Related models}. Another unsupervised learning algorithm based on NNs is the EncoderMap~\cite{Lemke2019}. This method combines an autoencoder with the cost function of Sketch-map~\cite{tribello2012using}. Sketch-map is a multidimensional scaling-like algorithm that aims to preserve the structural similarity, i.e. to reproduce in low-dimensional space the distances between points in the high-dimensional space. In \verb|mlcolvar|, this can be easily implemented by subclassing the AutoEncoder CV and adding the sketch-map objective to the loss function. 
In a similar spirit, also the Multiscale Reweighted Stochastic Embedding~\cite{rydzewski2021multiscale}, which combines a NN with the t-stochastic neighbor embedding (t-SNE) cost function can be implemented.

\subsection{Supervised methods}

\vspace{0.5em}
\paragraphtitle{Linear Discriminant Analysis (LDA)}. LDA~\cite{welling2005fisher} is a statistical analysis method that aims to find the best linear combination of input variables that maximally separates the given classes. 
This is achieved by maximizing the so-called Fisher's ratio which measures the ratio of between-class variance to the within-class one. Similarly to PCA, the discriminant components are found via the solution of a (generalized) eigenvalue problem involving the within and between-class covariance matrices. If for PCA the eigenvalues represent the variance, here they measure the amount of separation between states along the relevant eigenvectors. Note that for LDA the number of non-zero eigenvalues (and hence of CVs that can be used) is $C-1$ with $C$ being the number of metastable states. A variant, called harmonic-LDA (HLDA), has been employed for CVs design~\cite{Mendels2018,Piccini2018}. 

\vspace{0.5em}
\paragraphtitle{Neural-network based LDA (Deep-LDA)}.
A non-linear generalization of LDA can be obtained by transforming the input features via a NN~\cite{Dorfer2016,Bonati2020}, and then performing LDA on the NN outputs (see Table ~\ref{tab:methods}, panel b). In this way, we are transforming the input space in such a way that the discrimination between the states is maximal. During the training, the parameters are optimized to maximize the LDA eigenvalues (Fisher's loss). The CV(s) are then obtained by projecting the NN output features along the LDA eigenvectors. This has the advantage of obtaining orthogonal CVs. In the case of two states, maximizing the Fisher's loss is equal to maximizing the single eigenvalue, while in the multiclass scenario, we can either maximize their sum or just the smallest one~\cite{Dorfer2016}. Since the LDA objective is not bounded, a regularization must be added to avoid the projected representation from collapsing into delta-like functions, which would not be suitable for enhanced sampling applications~\cite{Bonati2020}.


\vspace{0.5em}
\paragraphtitle{Targeted Discriminant Analysis (Deep-TDA)}.
In Deep-TDA~\cite{trizio2021enhanced}, the discrimination criterion is achieved with a distribution regression procedure.
Here, the outputs of the NN are directly used as CVs (see Table ~\ref{tab:methods}, panel c), and the parameters are optimized to discriminate between the different metastable states. 
This is achieved by choosing a target distribution along the CVs equal to a mixture of Gaussians with diagonal covariances and preassigned positions and widths, one for each metastable state.
This targeted approach performs particularly well in the multi-state scenario, as it allows to exploit information about the dynamics of the system (i.e. a precise ordering of the states) to reduce further the dimensionality of the CVs space with respect to LDA-based methods.

\subsection{Time-informed methods}

\vspace{0.5em}
\paragraphtitle{Time-lagged Independent Component Analysis (TICA)}.
TICA~\cite{Naritomi2011,Perez-Hernandez2013} is a dimensionality reduction method that identifies orthogonal linear combinations of input features that are maximally autocorrelated, and thus represent the directions along which the system relaxes most slowly. For a given lag-time, these independent components are determined as the eigenfunctions of the autocorrelation matrix associated with the largest eigenvalues, which are connected to their relaxation timescales. These have been shown to approximate the eigenfunctions of the transfer operator~\cite{Prinz2011}, which is responsible for the evolution of the probability density toward the Boltzmann distribution. 
TICA has been applied both to enhanced sampling~\cite{Sultan2017a} and to extract the CVs from biased simulations~\cite{McCarty2017c,Yang2018}.

\vspace{0.5em}
\paragraphtitle{Neural network basis functions for TICA (Deep-TICA)}. 
Similarly to LDA and Deep-LDA, we can consider a nonlinear generalization of TICA by applying a NN to the inputs before projecting along the TICA components (see Table ~\ref{tab:methods}, panel b). This corresponds to using NNs as basis functions for the variational principle of the transfer operator~\cite{Prinz2011,Perez-Hernandez2013}.
Similar architectures have been proposed, which all aim to maximize the TICA eigenvalues (typically the sum of the squares is maximized)~\cite{Mardt2018,Chen2019,bonati2021deep}. The implementation in \verb|mlcolvar| follows the Deep-TICA~\cite{bonati2021deep} method. In addition, we implemented different ways of reweighting the data~\cite{McCarty2017c,yang2019enhanced,Chen2023} as well as a reduced-rank regression estimator~\cite{kostic2022learning} to learn more accurate eigenfunctions.

\vspace{0.5em}
\paragraphtitle{Time-lagged AutoEncoders}. 
Another class of methods that work with pairs of time-lagged data is based on autoencoding NNs. Time-lagged autoencoders (TAEs)~\cite{Wehmeyer2018b} have the same architecture as standard ones, but the encoder/decoder parameters are optimized to find a compressed representation capable of predicting the configuration after a given lag-time rather than reconstructing the inputs. Thus, the decoder takes the CV at time $t$ and uses it to reconstruct the time-lagged inputs at $t+\tau$.
In \verb|mlcolvar|, this can be simply achieved using an \verb|AutoEncoderCV| but with a dataset in which the output targets are time-lagged configurations.

Similarly, one can also consider a time-lagged variant of the variational autoencoder, as done in the Variational Dynamics Encoder (VDE) architecture~\cite{Hernandez2018}. To build a VDE, one needs to simply optimize a time-lagged VAE with an additional term in the loss function which maximizes the autocorrelation of the latent space.
It is worth noting that both TAEs and VDEs tend to learn a mixture of slow and maximum variance modes~\cite{Chen2019capabilities}, at variance with the non-linear generalizations of TICA which only learn slowly decorrelating modes.

\section{Examples}\label{sec:examples}
As discussed, the methods implemented in the library have been extensively applied to study a wide range of atomistic systems. In this section, we provide a didactic overview of the library's capabilities by focusing on a simple toy model. Specifically, we showcase an example for each of the CV categories presented above with the intent of highlighting the versatility of the implementation and how different workflows may be chosen depending on the available data.
For atomistic examples, we refer the reader to the documentation of the \texttt{mlcolvar} library, which includes notebooks demonstrating the use of the library with systems taken from the literature on data-driven CVs.

    \begin{figure}[h!]
        \centering
        \includegraphics[width=0.8\linewidth]{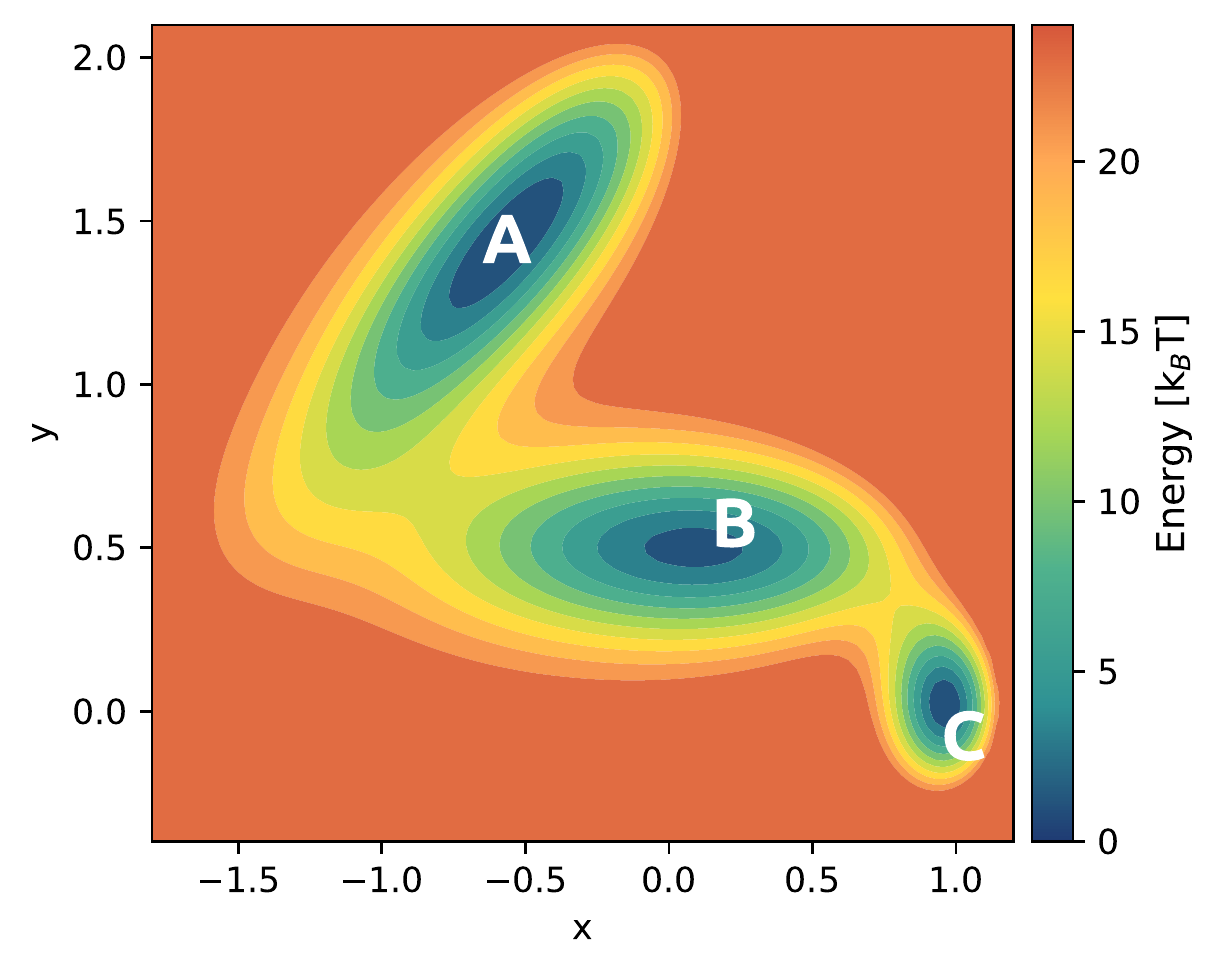}
        \caption{Three-state potential energy landscape for the movement of a particle in two dimensions, used as a toy model in sec.~\ref{sec:examples}. The analytic expression of the potential energy, obtained by modifying the Muller-Brown potential to have three states, is available on the GitHub repository in the Jupyter notebooks implementing the examples.}
        \label{fig:examples-potential}
    \end{figure}


In the following, we consider a particle moving in two dimensions under the action of the three-state potential depicted in Fig.~\ref{fig:examples-potential} built out of a  sum of Gaussians.
 the input features of the NNs are taken to be the $x$ and $y$ position of the particle. The activation function is chosen to be the shifted softplus~\cite{schutt2018schnet}, which is well suited for differentiating the CVs. The parameters are optimized via gradient descent using the ADAM optimizer with a learning rate of $10^{-3}$. The dataset is split into training and validation, and early stopping is used to avoid overfitting.
All the simulations are performed using the simple Langevin dynamics code contained in the \verb|ves| module of PLUMED, and the biased simulations are performed using the OPES~\cite{invernizzi2020rethinking} method with a pace of 500 steps, the automatic bandwidth selection, and a barrier parameter equal to 16 $k_B T$. 

\subsection{Unsupervised setting: state discovery}
We first start with the least informed scenario in which we only have data limited to a single metastable state and aim at exploring the potential energy surface. 
Starting from the first unbiased data in state A, we adopt an iterative procedure akin to the MESA~\cite{Chen2018} method in which we train an \verb|AutoEncoderCV|, perform a short biased run, and add the collected configurations to the training dataset. This workflow is repeated until necessary, e.g. all the states have been discovered. We performed 16 iterations of $10^5$ steps each and reported the results in Fig.~\ref{fig:examples_results}. In panel a, we colored the sampled regions according to the iteration in which they were first visited, while in panel b, we report the time evolution of the variable $y$, which is able to distinguish the different states. At first, the AE drives the sampling along the direction of maximum variance of state A (blue dots), but after a few iterations, it is able to discover state B as well. Finally, after 10 iterations, state C is also visited, and from there the system visits all three states, although only one or two transitions are observed per iteration. 

\subsection{Supervised setting: CVs as classifiers}\label{subsec:example_classifiers}
Once the three states of the system have been discovered, we can step up to more refined CV models based on supervised learning and aim for a comprehensive sampling of the free energy landscape, e.g. to converge a free energy profile.
For each of the three states, we run short unbiased MD runs and collect labeled samples of the three states (see Fig.~\ref{fig:examples_results}c). Then we train a \verb|DeepTDA| CV with a single component and a target distribution of three equidistant Gaussians (ordered as A, B, C for increasing CV values). This is motivated by the consideration that during the exploratory phase described above only transitions of the kind $A \leftrightarrow B$  and $B \leftrightarrow C$ are observed. 
Enhancing the sampling along the DeepTDA CV results in multiple transitions between the different states as reported in fig.~\ref{fig:examples_results}d. We observe that the sampling follows approximately the minimum free energy path connecting the states. Furthermore, the multiple transitions induced by this trial CV allow converging the free energy profile (reported in the inset of panel d).

\subsection{Time-lagged setting: improving CVs}
One scenario that often occurs in practice is when you have a suboptimal simulation, capable of promoting just a few transitions before getting stuck due to some slow orthogonal mode not being accelerated. 
In this context, time-informed methods such as \verb|DeepTICA| can be used to extract (approximations of) the slowly decorrelating modes that hamper simulations' convergence and thus design better CVs. 
As an example, we took a simulation performed using the position $y$ as the CV, which is the prototype of an order parameter capable of distinguishing the metastable states but not the transition states between them. As can be seen from fig.~\ref{fig:examples_results}e, this results in poor sampling. Following the DeepTICA scheme, we optimized a new CV based on this data, which, when biased, leads to a much more diffusive sampling capable of converging the simulation in a fraction of the time (fig.~\ref{fig:examples_results}f).

\subsection{Multitask learning: a semi-supervised application}
\begin{figure}[h!]
    \centering
    \includegraphics[width=\linewidth]{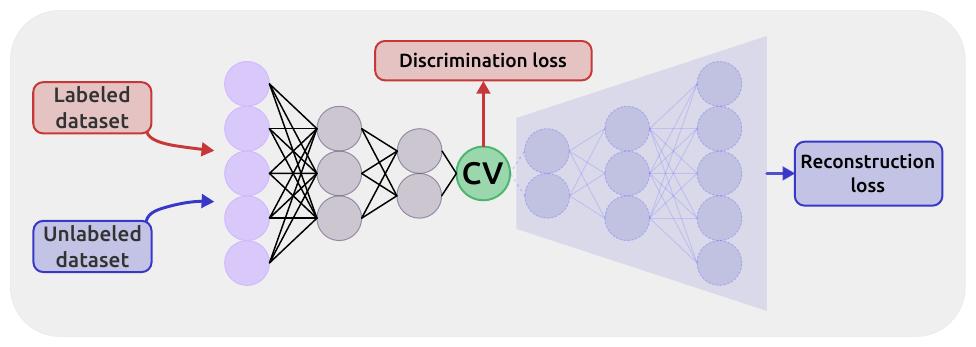}
    \caption{Example of a multitask CV employed for semi-supervised learning. 
    The model is based on an autoencoder architecture and is optimized on two separate datasets with different criteria.
    Data from an unlabeled dataset (blue path) contributes to an unsupervised loss function (i.e. reconstruction MSE loss)  which depends on the output of the decoder. Data from a labeled dataset (red path) contributes to a supervised loss function that depends directly on the CVs space, which is the output of the encoder (e.g. TDA loss). }
    \label{fig:multitask}
\end{figure}
\begin{figure*}
        \centering
        \includegraphics[width=0.76\textwidth]{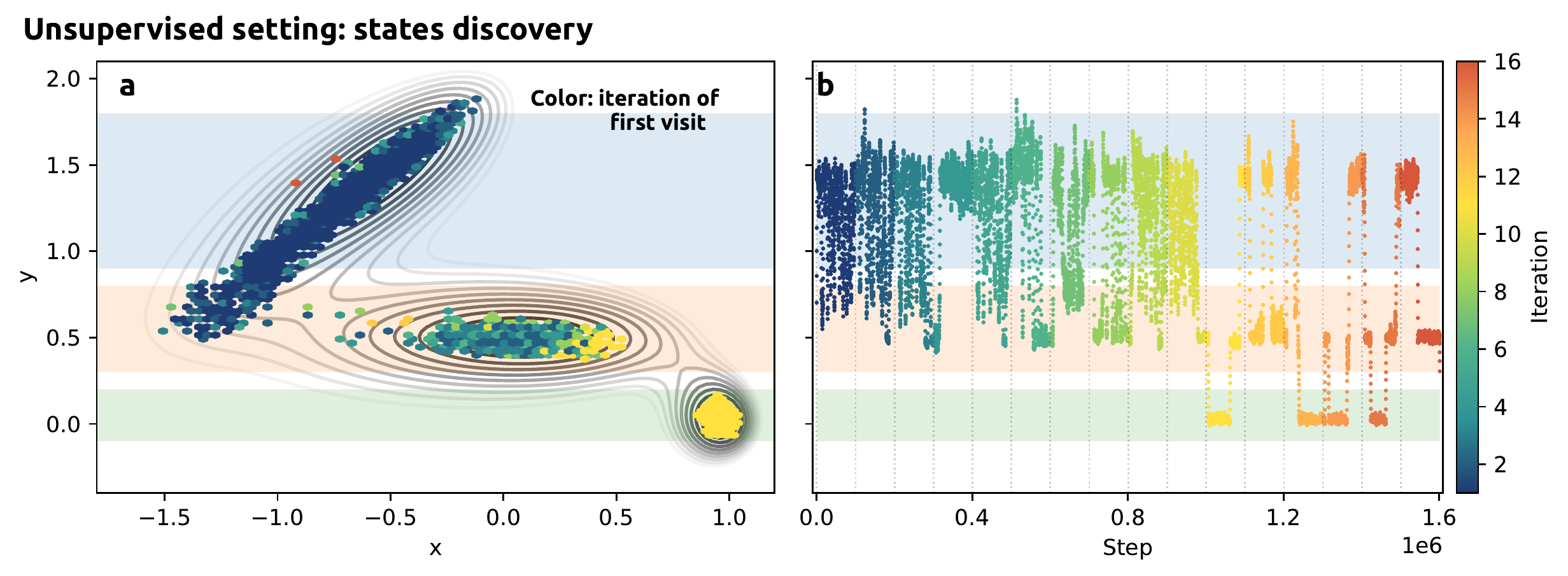}
        \includegraphics[width=0.76\textwidth]{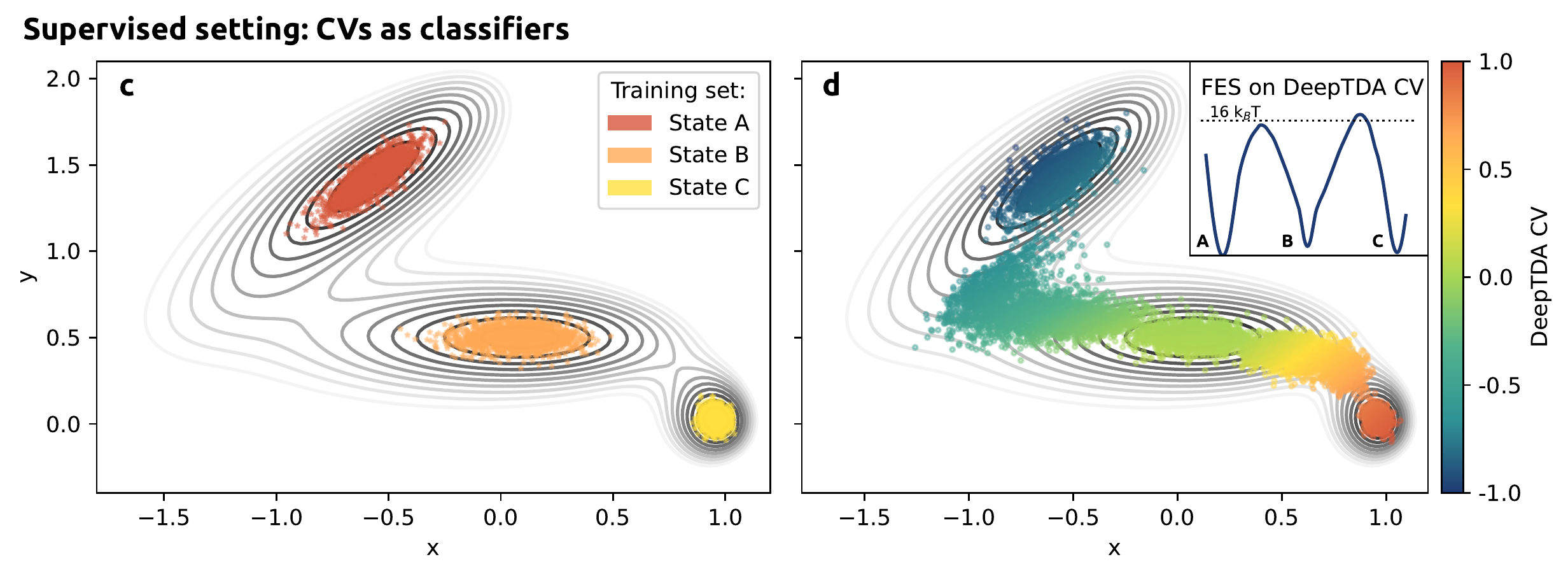}
        \includegraphics[width=0.76\textwidth]{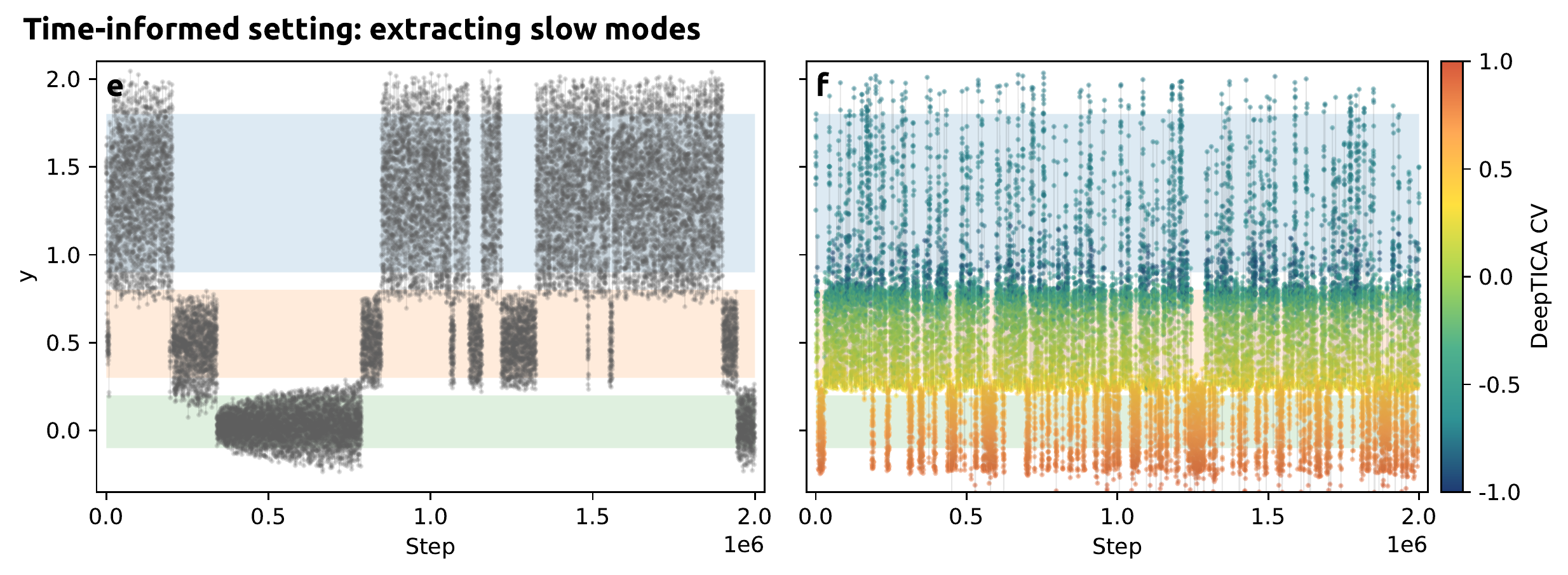}
        \includegraphics[width=0.76\textwidth]{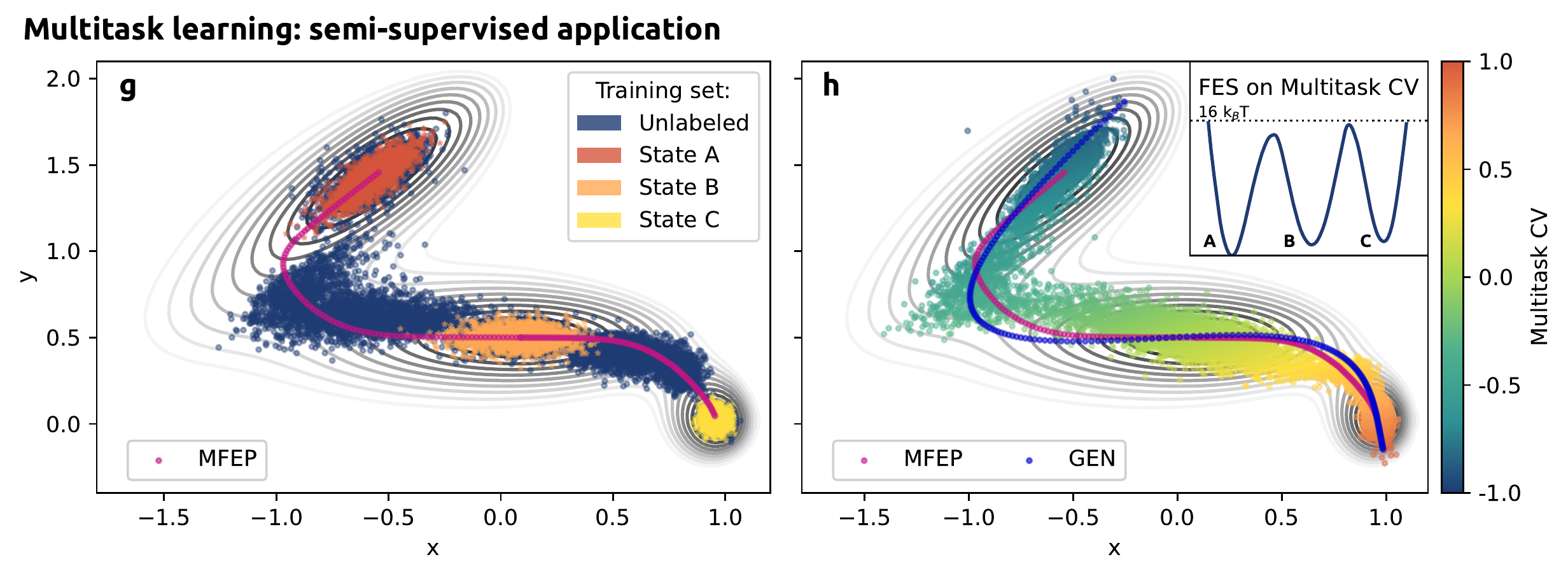}
        \caption{MD simulations of the toy model in Fig.~\ref{fig:examples-potential}. The scatter plots (panels a,c,d,g,h) in the $x,y$ plane are colored according to the quantities reported in the colorbars, while lines denote the isolines of the 2D potential energy. The time series (panels b,e,f) describe the time evolution of the $y$ coordinate, possibly colored according to the quantity in the corresponding colorbar, while the colored shaded regions in the background indicate the three metastable states. The insets in d,h report the free energy surface (FES) computed from the sampled data projected along the corresponding CV.
        \textbf{a}) Exploration of the energy landscape biasing the autoencoder CV. Each region of the space is colored according to the iteration in which it was visited for the first time. \textbf{b}) Time evolution of the $y$ coordinate along different iterations of the autoencoder CV. 
        \textbf{c}) Training set of the DeepTDA variable. 
        \textbf{d}) Points sampled biasing the DeepTDA CV.
        \text{e}) Suboptimal OPES simulation biasing the y coordinate, used as training set for the DeepTICA CV. 
        \textbf{f}) Time series of the simulation biasing the DeepTICA CV as well as the previous (static) bias. 
        \textbf{g}) Training set for the semi-supervised multitask CV (see Fig. \ref{fig:multitask}). 
        \textbf{h}) Points sampled biasing the MultiTask CV. The blue dotted path (GEN) is generated by applying the decoder to a collection of evenly spaced samples in the latent space. In g,e the purple path represents the minimum free energy path (MFEP).}
        \label{fig:examples_results}
    \end{figure*}

Finally, we show how the library can be used to combine different data in a multitask framework to improve data-driven CVs. 
To this end, we take as an example the two datasets from sec.~\ref{subsec:example_classifiers}. The first is the labeled dataset which has been used to construct the DeepTDA CV (fig.~\ref{fig:examples_results}c), while the second is composed by the unlabeled configurations generated by biasing the DeepTDA CV (fig.~\ref{fig:examples_results}d). We define a single multitask CV that, as schematically depicted in Fig.\ref{fig:multitask}, is composed of an autoencoder optimized upon two different losses: the first is the reconstruction loss on the unlabeled dataset (blue path in Fig.\ref{fig:multitask}), and the second is the TDA loss acting on the latent space optimized on the labeled dataset (red path Fig.\ref{fig:multitask}). In this way, one obtains a semi-supervised approach in which each component benefits from the other. Indeed, the autoencoder CV is regularized to distinguish the metastable states, and at the same time, the classifier CV is informed about regions outside the local minima.

As a result, when this multitask CV is employed to enhance sampling, it follows the minimum free energy path (purple dotted line in fig.~\ref{fig:examples_results}h) more closely than the simulation using only DeepTDA, which was exploring higher-energy pathways (fig.~\ref{fig:examples_results}g). Moreover, the multitask approach allows for inspecting the CV space by using the model in a generative way. To this end, we examine how a path connecting the states in the latent space is mapped by the decoder into the reconstructed input space. This generates a set of configurations shown in fig.~\ref{fig:examples_results}h that traces the free energy path between the three states remarkably well.

\section{Conclusions}

In this work, we presented \verb|mlcolvar|, a software library aimed at easily implementing, training, and using machine learning collective variables for enhanced sampling simulations.
The software comprises a Python framework built on PyTorch for the training of the CV and a contributed extension to the PLUMED software that enables their use in enhanced sampling simulations.
The library natively implements several methodologies for the data-driven identification of CVs available in the literature.
Furthermore, its modular structure facilitates the development of new methods and their cross-contamination, for instance, via multitask learning.

We believe this library can be a valuable tool to foster the contamination between the methods and spirit of machine learning with atomistic simulations. Moreover, this is not meant to be a static object but one that evolves and improves to better and better adapt to the needs of the enhanced sampling community, both in biophysical and chemical and physical fields. New methods and architectures may be added, but also features dedicated, for example, to interpreting machine-learned variables.



%
%

%

\begin{acknowledgments}
We are grateful to F. Mambretti, S. Perego, U. Raucci, D. Ray and S. Das for their feedback on the manuscript, and to P. Novelli and N. Pedrani for their contributions to the library.
L.B. acknowledges funding from the German Federal Ministry of Education and Research (BMBF) through the TransHyDE/Ammoref project.
A.R. acknowledges funding from the Helmholtz European Partnering program (``Innovative high-performance computing approaches for molecular neuromedicine'').\end{acknowledgments}

\section*{Code and Data Availability} \label{sec:code_avail}

The \verb|mlcolvar| package is freely available under the MIT license and can be downloaded from the GitHub repository \url{https://github.com/luigibonati/mlcolvar}. The library has been released on the Python Package Index (PyPI) for easy dissemination, and it is open to external contributions according to the guidelines provided in the library's documentation, which is available at \url{https://mlcolvar.readthedocs.io}. This includes a number of tutorials and examples in the form of Jupyter notebooks which can be automatically run in Google Colab.
The code to reproduce the experiments and to run enhanced sampling simulations is also available in the Jupyter notebooks provided in the documentation and has also been deposited in the PLUMED-NEST~\cite{plumed2019promoting} repository with \hyperlink{https://www.plumed-nest.org/eggs/23/022/}{plumID:23.022}.

\section*{Bibliography}
\bibliography{references}

\end{document}